\documentclass[floats,twocolumn,pra]{revtex4-2}
\usepackage{eurosym}
\usepackage{graphics}
\usepackage[thmmarks]{ntheorem}
\usepackage{graphicx}
\usepackage{algorithm}
\usepackage{algpseudocode}
\usepackage{amsmath}
\usepackage{amsfonts}
\usepackage{amssymb}
\usepackage{float}
\usepackage{longtable}
\usepackage{flushend}
\usepackage{epsfig}
\usepackage{latexsym}
\usepackage{theorem}
\usepackage{bbm}
\usepackage{bm}
\usepackage{psfrag}
\usepackage[utf8]{inputenc}
\usepackage[T1]{fontenc}
\usepackage{physics}%
\usepackage{color}
\usepackage{multirow}
\usepackage{float}
\usepackage{mathtools}
\usepackage{soul}
\usepackage{hyperref}
\usepackage{mathrsfs}
\usepackage[mathscr]{eucal}
\usepackage{ulem}
\usepackage[makeroom]{cancel}
\usepackage{placeins}

\DeclareMathOperator*{\argmin}{arg\,min}

\newcommand{\SP}[1]{{{\textcolor{black}{#1}}}}

\newcommand{\mr}[1]{{{\mathrm{#1}}}}

\begin{document}

\title{Fundamental Limits on QBER and Distance \\in Quantum Key Distribution}
\author{Stefano Pirandola}
\affiliation{Department of Computer Science, University of York, York YO10 5GH, United Kingdom}
 
\begin{abstract} 
Quantum key distribution (QKD) enables information-theoretic secure communication, yet its ultimate tolerance to noise and achievable transmission distance remain fundamentally constrained. We establish the maximum quantum bit error rate (QBER) compatible with secure QKD and derive corresponding upper bounds on communication distance. Our results follow from a fundamental capacity threshold for qubit Pauli channels and apply to protocols based on two or more mutually unbiased bases, using either single-photon or weak coherent sources. By connecting information-theoretic limits to realistic physical noise models, we obtain universal bounds on achievable distances in fiber and free-space links, including diffraction-limited constraints relevant to deep-space quantum communications. These findings clarify the ultimate noise robustness of QKD and delineate the fundamental boundaries of secure quantum communication.
\end{abstract}

\maketitle

\section{Introduction}
Quantum key distribution (QKD)~\cite{QKDrev1,BB84,six} is one of the most mature quantum technologies, currently being deployed in both research and industrial settings. Yet, QKD is affected by fundamental rate-loss limitations~\cite{RCI1,PLOB} which require the use of intermediate repeater stations to form chains or networks of nodes, which may be trusted or untrusted~\cite{TF1,TF2,TF3,TF4,TF5}.
The Pirandola-Laurenza-Ottaviani-Banchi (PLOB) bound~\cite{PLOB} establishes that $-\log_2(1-\eta)$ is the maximum possible number of key bits per channel use that can be extracted over a lossy channel with transmissivity $\eta$. While this provides the maximum key rate achievable in repeaterless QKD, it does not directly impose a bound on the maximum quantum bit error rate (QBER) that is tolerable by discrete-variable (DV) protocols. This is the main result of this paper.

We begin by establishing a necessary and sufficient condition under which all two-way assisted capacities of the qubit Pauli channel are strictly positive. Building on this result, we then derive the threshold QBER values required to guarantee the security of a DV-QKD protocol employing either two or three mutually unbiased bases (MUBs). Next, we adopt a standard noise model that relates the QBER to the channel transmissivity and other relevant system parameters. Exploiting this relationship, we translate the QBER threshold conditions into bounds on the maximum transmission distance achievable by DV-QKD protocols over fiber-based and free-space links. Finally, we extend our analysis from the repeaterless setting to repeater-assisted protocols.

Our investigation not only establishes the fundamental limits of noise robustness for DV-QKD, but also reveals gaps in the existing literature. In particular, we identify the possibility of undiscovered DV-QKD protocols capable of tolerating higher QBERs than those currently known. Furthermore, we show that diffraction-limited free-space implementations can achieve exceptionally long transmission distances, thereby supporting the feasibility of deep-space quantum communications.

\section{Quantum channels in DV-QKD}

In DV-QKD protocols, such as the BB84 protocol~\cite{BB84}, there are two quantum channels involved, one affecting the system carrier (e.g., a photon or a photon pulse) and the other affecting the degree of freedom that is used for encoding and decoding (e.g., the polarization qubit). The first quantum channel is a bosonic lossy channel that transmits photons with probability $\eta$, referred to as the channel transmissivity. The second channel is a qubit Pauli channel. The lossy channel determines the detection rate, while the Pauli channel sets the QBER. Together, they determine the final secret key rate.

A qubit Pauli channel $\mathcal{P}$ acts on a state $\rho$ as follows
\begin{align}
\mathcal{P}(\rho)
&=\sum_{k=0}^{3} p_k P_k \rho P_k^\dagger, 
\end{align}
where $\mathbf{p}:=\{p_k\}$ is a probability distribution associated with the Pauli operators $P_k\in\{I,X,Y,Z\}$. Explicitly,
\begin{equation}
X=\begin{pmatrix}0&1\\1&0\end{pmatrix},\quad
Y=\begin{pmatrix}0&-i\\ i&0\end{pmatrix},\quad
Z=\begin{pmatrix}1&0\\0&-1\end{pmatrix},
\end{equation}
with $I$ being the $2 \times2$ identity matrix. 

The Pauli error probabilities can be connected with the values of the QBERs in the encoding bases. If the security of the QKD protocol is based on two MUBs, such as $X$ and $Z$ of the BB84 protocol, we have that the corresponding QBERs, known as bit error rate $E_X$ and phase error rate $E_Z$, satisfy
\begin{equation}
  ~E_X= p_2+p_3,~E_{Z}= p_1+p_2.\label{E_p_relation}
\end{equation}
If the security of the QKD protocol is based on three MUBs, $X$, $Z$ and $Y$, as in the six-state protocol~\cite{six}, then the associated QBERs statisfy Eq.~\eqref{E_p_relation} and the additional relation $E_{Y}= p_1+p_3$. 

There are some relevant specific examples of Pauli channels. One is the bit-and-phase-flip (BPF) channel corresponding to the case with $p_2 =0$. Since this channel only induces $X$ and $Z$ errors, it is the simplest model compatible with QKD protocols based on two MUBs. Another relevant channel is the depolarizing one, which is a Pauli channel with equal errors in the three bases. This is typically parametrized as follows \begin{equation}
\mathbf{p}=\left\{1-\frac{3p}{4},\,\frac{p}{4},\,\frac{p}{4},\,\frac{p}{4}\right\}, \label{DEPchoice}
\end{equation} 
and its action on a qubit state \(\rho\) reduces to the following
\begin{align}
\mathcal{P}_{\text{depol}}(\rho)
&=(1-p)\rho+p\,\frac{I}{2}.
\end{align}

\section{Capacity Thresholds}\label{Sec:Capacities}
Ref.~\cite{PLOB} established the ultimate rates for entanglement distribution, quantum information transmission, and QKD in the absence of an intermediate relay. For specific channels, such as the bosonic lossy channel, the bosonic quantum-limited amplifier, and the dephasing and erasure channels, Ref.~\cite{PLOB} derived the exact analytical expressions of the two-way assisted entanglement distribution capacity $D_2$, the two-way assisted quantum capacity $Q_2$, and the two-way assisted secret-key capacity $K_2$, showing that these capacities coincide for these channels. Ref.~\cite{PLOB} also provided lower and upper bounds for the two-way assisted capacities of Pauli, amplitude-damping, and the thermal-loss channels. The latter channel was studied in several other papers, where the lower bound was improved~\cite{LB1,LB2,LB3,LB4,LB5}. 

Let us call $\mathcal{C}_2$ the generic two-way assisted capacity. Depending on the context, this can be $D_2$, $Q_2$ or $K_2$. As a direct consequence of Ref.~\cite{PLOB}, one obtains equivalent criteria for a quantum channel to satisfy $\mathcal{C}_2 = 0$. For a qubit Pauli channel, we show that 
\begin{equation}
    \mathcal{C}_2(\mathcal{P})=0 \iff p_{\max}:=\max\{p_k\} \le \tfrac{1}{2},\label{equivalence}
\end{equation}
and, for a depolarizing channel, we therefore have
\begin{equation}
    \mathcal{C}_2(\mathcal{P}_{\text{depol}})=0 \iff p \ge\tfrac{2}{3}. \label{equiDEPOL}
\end{equation}

The proof in Appendix~\ref{AppP} relies on the fact that the two-qubit Choi state of a qubit Pauli channel has non-positive partial transposition (NPT)~\cite{Peres,horo3_96}. Note that this simple proof cannot be extended to higher dimensions since, already for qutrits ($3 \times3$), there is bound entanglement. In Appendix~\ref{sec:qber_pauli_reduction}, we also justify the reduction to a Pauli channel, rather than considering a fully general qubit channel, by showing that it represents the worst-case quantum channel compatible with the observed matched-basis QBERs.

\section{Maximum tolerable QBER}
We now exploit Eq.~\eqref{equivalence} to establish the maximum tolerable QBER in repeaterless QKD. Let us start by considering QKD protocols that are based on two MUBs, with observed QBERs $E_X$ and $E_Z$. As the values of these QBERs gradually increase, the probability of the identity Pauli `error' $p_0=1-(E_X+E_Z)+p_2$ gradually decreases. Also, because the values of the QBERs increase from zero, we have that $p_0=p_{\mr{max}}$. According to Eq.~\eqref{equivalence}, we find that QKD is possible as long as $E_X+E_Z < 1/2+p_2$. In the worst-case scenario, Eve does not introduce $Y$ errors ($p_2=0$), so we have a BPF channel and the maximum QBER is bounded as follows
\begin{equation}
   E_X+E_Z < 1/2. \label{2sumE}
\end{equation}
If this inequality is satisfied, then there certainly exists a QKD protocol with a positive key rate. If this is violated, the key rate is certainly zero for any QKD protocol.

Assuming the symmetric condition $E_X=E_Z:=E$, we see that Eq.~\eqref{2sumE} becomes $E<1/4$. It is interesting to note that this threshold coincides with the QBER induced by an intercept-resend attack against the BB84 protocol. While it may be intuitive to understand that security is not possible above $1/4$, it is less obvious to recognize that a QKD protocol (with suitable data processing) exists with a positive key rate up to $E=1/4$. \SP{Previous studies~\cite{LO_max} have identified that BB84 can be made secure up to $E=18.9\%$ with suitable two-way data processing, and more recently up to $E=20\%$~\cite{Chau_max}. Thus, our result raises the question of identifying a more powerful mechanism (protocol and/or data processing) able to effectively outperform $E=20\%$ and approach the $1/4$ upper bound limit in the symmetric scenario. Note that this limit applies to single-qubit BB84-like protocols, where the QBER is tested directly on the transmitted qubits, and does not necessarily apply to encoded variants with filtering or post-selection~\cite{Wang_max}.} 

Now consider QKD protocols based on three MUBs, with observed QBERs $E_X$, $E_Z$ and $E_Y$. In this case, we repeat the previous reasoning to obtain $p_{\mr{max}}=p_0=1-(E_X+E_Z+E_Y)/2$. Therefore, Eq.~\eqref{equivalence} leads to  
\begin{equation}
    E_X+E_Z+E_Y<1.\label{3sumE}
\end{equation}
Assuming symmetry $E_X=E_Z=E_Y:=E$, we have $E<1/3$. No QKD protocol based on three MUBs can be secure above this value in symmetric conditions. In fact, one can show that this corresponds to the threshold for an intercept-resend attack. As before, the less obvious point is that this result guarantees the existence of a QKD protocol based on three MUBs able to achieve a positive key rate up to $E=1/3$. \SP{We know that the six-state protocol combined with two-way processing can achieve a QBER of 26.4\%~\cite{LO_max} and, more recently, the performance was extended to $27.6\%$~\cite{Chau_max} and $30\%$~\cite{Wang_max}. So it is another open question to discover a more powerful 3-MUB protocol (and/or data processing method) able to exceed $E=30\%$ and reach $1/3$. More precisely, the threshold $E<1/3$ can be approached by entanglement-distillation methods requiring a quantum memory~\cite{Wang_max}. It remains an open question whether the same limit can be saturated by a prepare-and-measure implementation.}

\section{Noise model for QBER in DV-QKD}
The QBER of a QKD protocol can be related to the total transmissivity
$\eta$ of the lossy channel connecting Alice and Bob. This relation
allows one to translate the maximum tolerable QBER into a maximum
achievable transmission distance. The exact translation depends on the noise model valid for the QBER, as we discuss below.

Consider the QBER $E$ associated with a generic MUB ($X$, $Z$ or $Y$). We can write the minimum value of $E$ compatible with the transmissivity $\eta$ and the relevant parameters of Alice's source and Bob's detectors. This `irreducible' QBER represents the inevitable error rate arising from loss and device
imperfections, and therefore lower-bounds any observed QBER. Within the standard adversarial framework of QKD, all such noise is conservatively attributed to Eve and must be treated as channel noise associated with her perturbation of the systems.

Assume that Bob's detector has dark count probability $Y_0$ and misalignment error probability $e_{\mr{det}}$, then the irreducible QBER has the form
\begin{equation}
    E=\frac{\gamma e_{\mr{det}}+(1-\gamma)Y_0/2}{\gamma+(1-\gamma)Y_0},\label{irreQBER}
\end{equation}
where the analytical form of $\gamma=\gamma(\eta)$ depends on Alice's source. For a $k$-photon source, we may write $\gamma(\eta)=1-(1-\eta)^k$, while for an attenuated coherent-state source with intensity $\mu$, we have $\gamma(\eta)=1 - e^{-\eta \mu}$. 

For a decoy-state QKD protocol~\cite{one_decoy_fnsz,decoy1,decoy2,decoy3,decoy4} with intensities $\mu_i$ and corresponding probabilities $q_i$, we can identify the intensity $\mu \in \{ \mu_{i}\}$ associated with the best-case QBER. This QBER takes the form $E_{\mu}$ given by the right-hand-side of Eq.~\eqref{irreQBER} with $\gamma(\eta)=1 - e^{-\eta \mu}$. Then, the overall QBER of the protocol $E$ is lower-bounded by $E_{\mu}$. See Appendix~\ref{app2} for more details on the QBER modelling.

\section{Maximum distance}
By combining the noise models of the previous section with the results on the maximum QBERs, we can bound the maximum distance achievable by QKD protocols under various assumptions. We consider the symmetric scenario where the QBERs associated with the different MUBs are equal to some common value $E$. This assumption can always be seen as a best-case scenario. In fact, if the QBERs differ, we can take their minimum $E=\min\{ E_Z,E_X,\ldots\}$ as a common value, thereby resulting in an upper bound on the maximum distance. Thus, our next results will provide distances beyond which QKD protocols are certainly not secure. 

For a 2-MUB protocol, we can combine Eq.~\eqref{irreQBER} with the symmetric threshold $E<1/4$ to get the bound
\begin{equation}
    \gamma(\eta)>\Gamma_2:=\frac{Y_0}{1+Y_0-4 e_{\mr{det}}}\simeq Y_0.\label{maxeta}
\end{equation}
Then, for a 3-MUB protocol, the previous formula in Eq.~\eqref{maxeta} is slightly modified into the following
\begin{equation}
    \gamma(\eta)>\Gamma_3:=\frac{Y_0}{2+Y_0-6 e_{\mr{det}}}\simeq Y_0/2.\label{maxeta2}
\end{equation}

Let us set $\Gamma=\Gamma_2~\mr{or}~\Gamma_3$ depending if we consider 2- or 3-MUB protocols. Then, for a single-photon implementation ($\gamma=\eta$), the bounds can compactly be written as $\eta>\Gamma$. For a practical implementation with an attenuated source of intensity $\mu$ (or a decoy-state implementation with best-case intensity $\mu$), we instead have 
\begin{equation}
    \eta > -\frac{1}{\mu} \ln \left( 1-\Gamma\right).
\end{equation} 

The previous lower bounds for the transmissivity $\eta$ can be written in terms of upper bounds for the distance $d$ between Alice and Bob. First, we notice that we may decompose $\eta = \eta_{\mr{eff}}\eta_{\mr{ch}}$, where $\eta_{\mr{eff}}$ is the setup efficiency, i.e., the product of the transmissivities of Alice's and Bob's setups, while $\eta_{\mr{ch}}$ is the actual channel transmissivity, depending on the distance. In a fiber-based implementation, the fiber distance $d$ can be computed assuming a loss-rate $\alpha$ in dB/km, using $\eta_{\mr{ch}}(d) = 10^{-\alpha d/10}$. Therefore, by using $\gamma =\gamma[ \eta_{\mr{eff}}\eta_{\mr{ch}}(d)]$ in Eqs.~\eqref{maxeta} or~\eqref{maxeta2}, we can bound the maximum distance achievable in fiber by repeaterless DV-QKD protocols. We may write
\begin{equation}
d < -\frac{10}{\alpha}\log_{10}\Omega,~~
\Omega :=
\begin{cases}
\Gamma/\eta_{\mr{eff}} & \text{1-photon}, \\
\frac{-\ln(1-\Gamma)}{\eta_{\mr{eff}}\mu} & \text{attenuated}.
\end{cases}\label{omega}
\end{equation}

We can now derive numerical results. We assume \SP{high efficiency $\eta_{\mr{eff}}=0.9$}, low-loss fiber $\alpha=0.17$~dB/km, low dark counts $10^{-8}$ (compatible with SNSPD detectors) and low misalignment error $e_{\mr{det}}=1\%$. We find the results in Table~\ref{tabkm}, which represent distances beyond which no key can be extracted (\SP{for the chosen parameters}). Note that the current distance records for repeaterless QKD experiments are not far from these values~\cite{Long1}. 
\begin{table}[h]
\centering
\begin{tabular}{|c|c|c|c|}
\hline
\multicolumn{2}{|c|}{2-MUB} & \multicolumn{2}{c|}{3-MUB} \\ \hline
1 photon  & $\mu=0.3$  & 1 photon & $\mu=0.3$  \\ \hline
\SP{$466$~km}  & \SP{$436$~km}  & \SP{$484$~km} &  \SP{$454$~km}  \\ \hline
\end{tabular}
\caption{Examples of maximum fiber-distance for various protocols (see text for the parameters used).}
\label{tabkm}
\end{table}

\begin{figure}[t]
\vspace{0.2cm}
\centering
\includegraphics[width=0.45\textwidth]{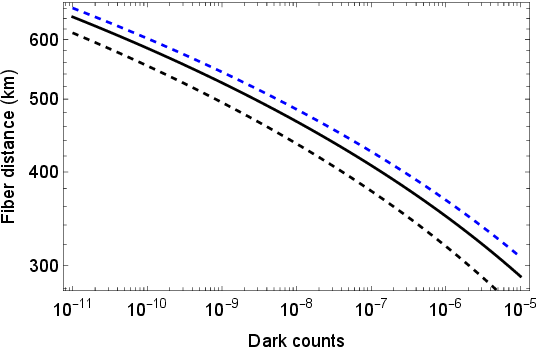}
\caption{Maximum fiber-distance in km versus dark count probability for a 2-MUB protocol with a single-photon source (solid black line), an attenuated source with intensity $\mu=0.3$ (dashed black line) and intensity $\mu=2$ (dashed blue line). We assume \SP{$\eta_{\mr{eff}}=0.9$}, $\alpha=0.17$~dB/km and $e_{\mr{det}}=1\%$.}
\label{fig:maxplot}
\end{figure}

It is interesting to note how the maximum distance depends on the dark counts, as shown in Fig.~\ref{fig:maxplot}. Because $Y_0 \ll 1$, we have the approximations in Eqs.~\eqref{maxeta} or~\eqref{maxeta2}. For single-photon protocols, this means $d \simeq -\log_{10}Y_0$.  

\section{Maximum free-space distance}
In free space, optical links are more generally described
as fading channels, since pointing errors and large-scale turbulence induce random beam wandering across the receiver aperture~\cite{freespace,PirandolaSAT,Trusted}. For our upper bound, we ignore the fading process and consider the maximum transmissivity of the free-space link. Its value is determined by diffraction, small-scale turbulence, and atmospheric
absorption. In general, we may decompose $\eta_{\mr{ch}}=\eta_{\mr{fs}}\eta_{\mr{atm}}$. 

On the ground, the atmospheric component can be modelled
by the Beer-Lambert extinction equation
\begin{equation}
\eta_{\text{atm}}(d)=\exp[-\alpha(h)d], \label{BLmain}%
\end{equation}
where $d$ is the horizontal path length in the atmosphere, and $\alpha(h)$ is
the extinction factor at the altitude $h$~\cite{Huffman,Hemani,Duntley}. A good approximation valid for the optical wavelength of $\lambda=800$~nm at near sea-level is $\alpha(h)=\alpha_{0}\exp(-h/H)$, where $\alpha_{0}\simeq5\times10^{-3}$ km$^{-1}$ and $H=6.6~$km~\cite{Vasy19,freespace}. 
For a satellite link, the atmospheric extinction varies with the altitude. For a satellite at slant distance $d=d(h,\theta)$, where $h$ is the altitude and $\theta \le 1$ is the zenith angle, we can write $\eta_{\mr{atm}}(d)\simeq [\eta_{\mr{atm}}^{\mr{zen}}(h)]^{\sec(\theta)}$, where $\eta_{\mr{atm}}^{\mr{zen}}$ is the transmissivity at the zenith position. For $h>30$~km, we can approximate $\eta_{\mr{atm}}^{\mr{zen}} \simeq 0.967$ at $\lambda=800$~nm~\cite{PirandolaSAT}.

The free-space component $\eta_{\text{fs}}$ is associated with beam waist broadening induced by diffraction and small-scale turbulence (inner-scale turbulence effects). Approximate analytical formulas can be written for weak and strong regimes of turbulence, while the intermediate behaviour can only be explored numerically~\cite{Yura73,Fante75}. Starting from the adopted turbulence model, one can write or numerically evaluate the spot size of the beam at distance $d$ and compute $\eta_{\text{fs}}(d)$. Then, by using $\gamma =\gamma[ \eta_{\mr{eff}}\eta_{\mr{fs}}(d)\eta_{\mr{atm}}(d)]$ in Eqs.~\eqref{maxeta} or~\eqref{maxeta2}, we can bound the maximum distance for free-space QKD, which can be applied to ground-based or ground-satellite implementations. 

\SP{Here, we derive an upper bound which only accounts for diffraction (extension to turbulence is discussed in Appendix~\ref{app:turbulence}). This bound applies to all free-space implementations, but is specifically meaningful for deep-space communications.} This diffraction-limited bound can be derived by using $\eta_{\mr{atm}}(d) \le 1$ and
\begin{equation}
    \eta_{\text{fs}}(d) \le \eta_{\text{diff}}(d)=1-e^{-2a_{R}^{2}/w_d^{2}} \simeq\frac{2a_R^2}{w_d^{2}}, \label{etadMAIN}
\end{equation}
where $a_R$ is the radius of the receiver circular aperture, and $w_d$ is the beam spot size at distance $d$. We have
\begin{equation}
 w_d^{2}=w_{0}^{2}\left[  \left(  1-d/R_{0}\right)  ^{2}+\left(
d/d_{R}\right)  ^{2}\right] \ge w_{0}^{2}\left(
d/d_{R}\right)^2,\label{waistdiff}
\end{equation}
where $w_{0}$ is the field spot size of the Gaussian beam generated by Alice, with carrier wavelength $\lambda$ and curvature $R_{0}$, while  $d_{R}:=\pi w_{0}^{2}\lambda^{-1}$ is the Rayleigh range~\cite{Siegman,svelto,Andrews93,Andrews94}. In this way, we get the following diffraction-limited upper bound for the maximum QKD distance
\begin{equation}
d < \frac{\pi w_0 a_R}{\lambda}\sqrt{\frac{2}{\Omega'}},~~
\Omega' := -\ln(1-\Omega)\simeq \Omega,
\end{equation}
where $\Omega$ takes the expressions in Eq.~\eqref{omega}, depending on the use of single-photon or attenuated QKD sources.

\SP{Let us investigate the limits induced by this bound assuming $\eta_{\mr{eff}}=0.9$, $Y_0=10^{-8}$ and $e_{\mr{det}}=1\%$ as before. Consider Alice sending a collimated Gaussian beam (\SP{$R_0 = +\infty$}) of wavelength $\lambda=800~$nm and waist $w_{0}=30$~cm, and Bob using an aperture of radius $a_{R}=10~$cm. Adopting a single-photon 2-MUB protocol, we get $d<1.54 \times 10^6$~km, compatible with the distance of the James Webb Space Telescope (at the Sun-Earth L2 Lagrange point). As discussed in Appendix~\ref{app:turbulence}, this diffraction-limited bound is also an excellent approximation in the presence of turbulence when we consider the downlink. By contrast, in uplink, the effect of turbulence is non-negligible, and the distance is reduced by one order of magnitude.}

\section{Repeater-based protocols}
In order to extend the results to repeater-based protocols, we can leverage Ref.~\cite{repeaterCAP}. Let us assume that Alice and Bob are connected by $N$ repeaters, so we have $N+1$ links. Each link $i$ induces a lossy channel and a Pauli channel $\mathcal{P}_i$. For each $\mathcal{P}_i$ with probabilities $\{ p_k^i\}$, we consider $p_{\mr{max}}^i:=\max_k{p_k^i}$. For the entire sequence of Pauli channels $\{ \mathcal{P}_i\}$, we take the bottleneck value $p_{\max}^{\min}:=\min_i p_{\max}^i$. As briefly explained in Appendix~\ref{app:rep},
the end-to-end two-way assisted capacity of the repeater chain $\mathcal{C}_2(\{\mathcal{P}_i\})$ satisfies
\begin{equation}
    p_{\max}^{\min} \le 1/2 \implies \mathcal{C}_2(\{\mathcal{P}_i\})=0.\label{repimplication}
\end{equation}

As before, this condition can be translated into a threshold for the maximum QBER. If each link on the chain implements a 2-MUB protocol with QBERs $E_Z^i$ and $E_X^i$, we must have 
\begin{equation}
    \max_i\{E_X^i+E_Z^i\}<1/2.
\end{equation}
If this threshold is violated, then no key rate can be extracted. However, since the converse implication in Eq.~\eqref{repimplication} does not hold, we cannot assert the existence of a repeater-based protocol that achieves a positive key rate when the total QBER approaches $1/2$. An analogous situation arises if the chain implements 3-MUB protocols, in which case the corresponding bound would be written as $\max_i\{E_X^i+E_Z^i+E_Y^i\}<1.$

Finally, these results admit a straightforward interpretation in terms of maximum transmission distance. In particular, the distance constraint identified for the repeaterless configuration also governs the longest individual segment within a repeater chain. 

\section{Conclusions}
In this work, we established the maximum tolerable QBER for DV-QKD protocols and bounded their maximum distance in both fiber-based and free-space implementations. We considered protocols based on two or more MUBs, and whose implementation employs either single-photon or weak coherent (attenuated) sources. While we derived the main results for the repeaterless configuration, we also showed that their extension to the repeater-based setting is immediate.

Our results imply the existence of undiscovered \SP{BB84-like} protocols and/or data processing methods that are able to overcome the best performance known so far. Identifying such protocols and/or methods is left as an open question for the community. In the free-space scenario, we showed that geometric diffraction imposes a fundamental upper bound on the maximum transmission distance; nevertheless, this limit remains sufficiently large to enable long-range implementations, including deep-space quantum communications.   

\bigskip

\noindent \textbf{Acknowledgments.} This work was supported by the Integrated Quantum Networks (IQN) Research Hub (EPSRC, Grant No. EP/Z533208/1).


\FloatBarrier

\appendix

\section{Zero-capacity thresholds}\label{AppP}
This is a simple proof of Eq.~(6) of the main text. First observe that the two-way assisted capacity of a Pauli channel $\mathcal{P}$ with probabilities $\{ p_k\}$ satisfies~\cite{PLOB}
\begin{align}
&\mathcal{C}_2(\mathcal{P}) \le \Phi(\mathcal{P}):= 
\begin{cases}
1-H_2(p_{\max}), & p_{\max}\ge \tfrac12,\\
0, & \text{otherwise},
\end{cases}
\label{strettoPAULI}
\end{align}
where $H_2$ is the binary Shannon entropy computed on $p_{\mr{max}}:=\max_k p_k $. From Eq.~\eqref{strettoPAULI}, we see that $p_{\max} \le \tfrac12$ implies $\mathcal{C}_2(\mathcal{P})=0$. Then we ask if \(\mathcal{C}_2(\mathcal{P})=0\)
implies \(p_{\max} \le \tfrac{1}{2}\), so we can write an equivalence. 

A basic strategy for distributing entanglement over a Pauli channel $\mathcal{P}$ is to send half of a Bell state through the channel. Alice keeps one qubit and sends the other qubit to Bob. After the channel, the joint two-qubit output is the Choi state of the qubit Pauli channel. This is a Bell-diagonal  state with spectral decomposition
\begin{equation}
\rho_{\mathcal{P}}=\sum_{k=0}^{3} p_k\,\Phi_k, \label{Choi_out}
\end{equation}
where the eigenvalues coincide with the channel probabilities and
$\{\Phi_k\}$ are the four Bell states.

 It is immediate to check that the two-qubit Choi state $\rho_{\mathcal{P}}$ in Eq.~\eqref{Choi_out} is an NPT state~\cite{Peres,horo3_96} for  $p_{\max} > 1/2$. Because, in dimensions $2 \times 2$, all NPT states have non-zero distillable entanglement
 , we have that $\rho_{\mathcal{P}}$ has non-zero distillable entanglement $E_{D}(\rho_{\mathcal{P}})>0$. Thus, this basic strategy for entanglement distribution guarantees that entanglement can be distilled from the output state of the Pauli channel if we have $p_{\max} > 1/2$. This means that $p_{\max} > 1/2$, implies $\mathcal{C}_2(\mathcal{P}) \ge E_{D}(\rho_{\mathcal{P}}) >0$ or, equivalently, that $\mathcal{C}_2(\mathcal{P})=0$ implies $p_{\max} \le 1/2$.
 
 Finally, because the depolarizing channel is a Pauli channel with probabilities as in Eq.~(4) of the main text, we automatically derive Eq.~(7) of the main text.

\section{Reduction to Pauli Channel}\label{sec:qber_pauli_reduction}
Here we clarify the sense in which the Pauli-channel model used in the main text represents the worst-case channel compatible with the observed QBERs. The analysis is not meant to reconstruct an arbitrary qubit channel from the experimental data. Rather, it follows the standard QBER-only approach of DV-QKD, where Alice and Bob estimate the error probabilities in the relevant mutually unbiased bases and then consider the most conservative effective channel compatible with these observations.

Let $\mathcal{N}$ be a general qubit channel compatible with a given set of matched-basis QBERs. In a Pauli-transfer-matrix representation, the matched-basis QBERs determine the diagonal correlations, whereas cross-basis correlations correspond to off-diagonal terms. Applying the standard Pauli twirl,
\begin{equation}
    \mathcal{T}(\mathcal{N})(\rho)
    :=
    \frac{1}{4}
    \sum_{P\in\{I,X,Y,Z\}}
    P\,\mathcal{N}(P\rho P)\,P ,
\end{equation}
removes the off-diagonal terms and produces a Pauli channel $ \mathcal{P}(\rho)$.
Moreover, since the twirl is based on local operations and classical communication (LOCC), it cannot increase the two-way assisted capacity. Therefore,
\begin{equation}
    C_2(\mathcal{P})
    =
    C_2[\mathcal{T}(\mathcal{N})]
    \leq
    C_2(\mathcal{N}) .\label{Cmono}
\end{equation}

This monotonicity property is well known for the two-way assisted entanglement-distribution capacity $D_2$, and therefore also for the two-way assisted quantum capacity $Q_2$. Indeed, both quantities are monotonic under LOCCs. The same reasoning also applies to QKD, namely to the two-way assisted secret-key capacity $K_2$. In fact, any secret-key protocol designed for $\mathcal{T}(\mathcal{N})$ can be simulated over the original channel $\mathcal{N}$ as follows: Alice and Bob first implement the same random Pauli pre- and post-processing that defines the twirl, and then run the secret-key protocol for $\mathcal{T}(\mathcal{N})$. Since these additional operations are allowed in a two-way assisted QKD protocol, the original channel cannot have a smaller secret-key capacity than its twirled version. Hence,
\begin{equation}
    K_2[\mathcal{T}(\mathcal{N})] \leq K_2(\mathcal{N}) .
\end{equation}
Thus, Pauli twirling cannot increase any of the relevant two-way assisted capacities, including $D_2$, $Q_2$, and $K_2$.

Because of Eq.~\eqref{Cmono}, the Pauli channel compatible with the observed matched-basis errors is the least favourable representative of the corresponding QBER class: any additional cross-basis correlations present in a more general qubit channel can only increase the two-way assisted capacity relative to this Pauli representative. This is exactly the conservative viewpoint adopted in a QKD security analysis based solely on the observed QBERs. Since Alice and Bob do not use information beyond the matched-basis error rates, they must consider the least favourable channel compatible with those observations. The Pauli representative is therefore the worst-case effective channel used throughout our analysis.

\section{QBER noise modelling}\label{app2}
Let us write the details of relevant formulas here. Assume that Alice sends a $k$-photon state through a lossy channel with transmissivity $\eta$, and Bob has a threshold detector at its output, which is the standard case in DV-QKD. Then, the transmissivity of a $k$-photon signal is
\begin{equation}
    \eta_k = 1-(1-\eta)^k,\label{eta_k}
\end{equation}
so that $\eta_0 =0$, $\eta_1=\eta$, etc. This is the probability that at least one photon passes the threshold detector.

Given an input $k$-photon state, we adopt the union model in which a detection
event occurs either due to a signal click or due to a dark count in the absence of a signal click. For this reason, the yield takes the form
\begin{equation}
Y_k=\eta_k+(1-\eta_k)Y_0,
\end{equation}
where $Y_0$ is the dark count probability. Correspondingly, we also consider the error yield, i.e., the overall probability of an erroneous detection. This is given by
\begin{equation}
P_k = e_{\mathrm{det}}\eta_k +(1-\eta_k)Y_0/2,
\end{equation}
where $e_{\mathrm{det}}$ is the misalignment error associated with
signal clicks. Thus, the minimum QBER reads
\begin{equation}
E_k=\frac{P_k}{Y_k}
\simeq\frac{e_{\mathrm{det}}\eta_k+Y_0/2}{\eta_k+Y_0}.\label{kQBER}
\end{equation}
In particular, if Alice has an ideal single-photon source, then the QBER is given by Eq.~\eqref{kQBER} for $k=1$.

In practical QKD implementations, Alice has an attenuated coherent-state source with mean number of photons (intensity) $\mu$, which generates $k$-photon signals according to a Poisson distribution $p^{\mu}_A(k)=(\mu^k/k!)e^{-\mu}$.
For the intensity $\mu$, we consider the gain $Q_{\mu}$, which is the overall probability of a detection event, and the error gain $P_{\mu}$, which is the overall probability of an error
. These are given by
\begin{align}
    P_{\mu}&=e_{\mr{det}} (1 - e^{-\eta \mu})+e^{-\eta \mu}Y_0/2 , \\
    Q_{\mu} &= 1 - e^{-\eta \mu} + e^{-\eta \mu}Y_0.
\end{align}
The QBER associated with the intensity $\mu$ is
\begin{equation}
    E_{\mu} = \frac{P_{\mu}}{Q_{\mu}} \simeq \frac{e_{\mr{det}} (1 - e^{-\eta \mu})+Y_0/2 }{1 - e^{-\eta \mu} + Y_0}.\label{QBERformula} 
\end{equation}  

In the case of decoy-state protocols, Alice randomly chooses between various intensities $\mu_i \ge0$ with probabilities $q_i>0$. The expected minimum QBER can be written as   
\begin{equation}
    E = \frac{ \sum_i  c_{i} q_i P_{\mu_i}} {\sum_i  c_{i} q_i Q_{\mu_i}},\label{expected_QBER}
\end{equation}
where $c_{i}=[1+r_s Q_{\mu_i} \tau_{\mr{dt}}]^{-1}$ is a factor depending on the repetition rate of Alice's source $r_s$ and the dead time $\tau_{\mr{dt}}$ of Bob's detector~\cite{one_decoy_fnsz}.

Setting $E_{\mu_i}=P_{\mu_i}/Q_{\mu_i}$, it is easy to check that 
\begin{equation}
    \min_i\{E_{\mu_i}\} \le E  \le \max_i \{E_{\mu_i}\} \le \sum_i E_{\mu_i}.\label{convex}
\end{equation} 
In fact, we can assume $Q_{\mu_i} > 0$ for all $i$, and write
\begin{align}
E 
= \sum_i \lambda_i\, E_{\mu_i},
\end{align}
where
\begin{equation}
\lambda_i := \frac{c_i q_iQ_{\mu_i}}{\sum_j c_j q_j Q_{\mu_j}},~\lambda_i > 0,~ \sum_i \lambda_i = 1.
\end{equation}
Because $E$ is convex in $E_{\mu_i}$, we have Eq.~\eqref{convex}.

As a result, we can choose $i^*=\argmin_i E_{\mu_i}$ and consider $E_{\mu_{i^*}}$ as the minimum achievable QBER. This is equivalent to use $E_{\mu}$ in Eq.~\eqref{QBERformula} where we set $\mu=\mu_{i^*}$. In our study, we therefore use Eq.~\eqref{QBERformula} as the minimum achievable QBER for decoy-based QKD implementations. It is understood that this provides a best-case scenario, so that using this modelling will provide an upper bound for the maximum distance.


\section{Including turbulence\label{app:turbulence}}
Let us provide analytical estimates for the free-space component $\eta_{\mr{fs}}$ in the presence of turbulence. In general, a Gaussian beam will be affected by both beam spreading (due to diffraction and fast time-scale turbulence) and beam wandering (due to slow time-scale turbulence). One can define the long-term spot size of the beam 
\begin{equation}
w_{\text{lt}}^{2}=w_{\text{st}}^{2}+\sigma_{\text{TB}}^{2},\label{ltspot}
\end{equation}
where $w_{\text{st}}$ is the short-term spot size (larger than the diffraction-limited spot size and induced by the fast dynamics of the smaller turbulent eddies), while $\sigma^2_{\text{TB}}$ is the variance of the beam wandering (induced by the slow dynamics of the  large turbulence eddies). 

Let us consider a ground-satellite link at the optimal zenith angle ($\theta = 0$), which can be a downlink or an uplink. The ground station is at sea level ($0$ meter altitude) and the satellite is at altitude $h$ expressed in meters. We use the Hufnagel-Valley (H-V) model of atmospheric turbulence~\cite{Stanley,Valley}. More precisely, we use the H-V$_{5/7}$ model~\cite[Sec.~12.2.1]{AndrewsBook}, which is appropriate for nighttime and low winds. In this model, the structure constant takes the following form
\begin{align}
C_{n}^{2}(h)  &  =5.94\times10^{-53}\left(  \frac{v}{27}\right)  ^{2}%
h^{10}e^{-h/1000}\nonumber\\
&  +2.7\times10^{-16}e^{-h/1500}+Ae^{-h/100}, \label{HFmodel}%
\end{align}
where $v=21~$m/s and $A=1.7\times10^{-14}~$m$^{-2/3}$. Studying the Rytov variance~\cite{RytovAPPROX}, one can check that the regime of weak turbulence is well satisfied for both downlink and uplink, as discussed in Appendix C.3 of Ref.~\cite{PirandolaSAT}. 

Starting from the structure constant and the wavenumber $k=2\pi/\lambda$ (with $\lambda$ being the wavelength of the beam), one can introduce the spherical-wave coherence length $\rho_{0}$ for uplink ($\uparrow$) and downlink ($\downarrow$). At the zenith, they have these expressions
\begin{align}
&  \rho_{0}^{\uparrow/\downarrow}=\left[  1.46k^{2}\int_{0}^{h}d\xi\left(
1-\tfrac{\xi}{h}\right)  ^{\frac{5}{3}}\gamma^{\uparrow/\downarrow}(\xi)\right]
^{-\frac{3}{5}},\nonumber\\
&  \gamma^{\uparrow}(\xi)=C_{n}^{2}(\xi),~\gamma^{\downarrow}(\xi)=C_{n}^{2}(h-\xi).
\end{align}

Using the spherical-wave coherence length, we can approximate the long-term spot size. In particular, for downlink, we have~\cite{Fante75}
\begin{equation}
w_{\text{lt}}^{2}\simeq w_{h}^{2}+2\left( \frac{\lambda h}{\pi\rho
_{0}^{\downarrow}} \right)  ^{2},
\end{equation}
where $w_h$ is the diffraction-limited beam spot size at altitude $h$. From the value of $\rho
_{0}^{\downarrow}$, one can check that, already at the Kármán line ($h=10^{5}$~m) and for typical optical frequencies ($\lambda=800$~nm), diffraction
is the only relevant effect, i.e., $w_{\text{lt}}\simeq
w_{h}$. In this case, we can reduce the analysis to the diffraction-limited bound of the main text, which provides the results in Table~\ref{tabDOWN}.
\begin{table}[h]
\centering
\begin{tabular}{|c|c|c|c|}
\hline
\multicolumn{2}{|c|}{2-MUB} & \multicolumn{2}{c|}{3-MUB} \\ \hline
1 photon  & $\mu=0.3$  & 1 photon & $\mu=0.3$  \\ \hline
$1.54 \times 10^6$ & $0.84 \times 10^6$  & $2.20 \times 10^6$ &  $1.20 \times 10^6$  \\ \hline
\end{tabular}
\caption{Bounds on the downlink distance in km, for the various protocols. These coincide with the diffraction-limited bounds computed using Eq.~(18) of the main text. Parameters are: collimated beam ($R_0=+\infty$) with $\lambda=800$~nm and initial beam waist $w_0=30$~cm; receiving aperture with radius $a_R=10$~cm; detector with efficiency $\eta_{\mr{eff}}=0.9$, dark counts $Y_0=10^{-8}$ and misalignment error $e_{\mr{det}}=1\%$.}
\label{tabDOWN}
\end{table}

By contrast, in uplink, the value of $\rho_{0}^{\uparrow}$ becomes too small (e.g., in the order of $1$~cm at $\lambda=800$~nm) and we can adopt the planar approximation~\cite[App.~C.4]{PirandolaSAT} 
\begin{align}
\rho_{0}^{\uparrow} &  \simeq\rho_{p}^{\uparrow}\simeq\left[  1.46k^{2} \int_{0}^{\infty}d\xi C_{n}^{2}(\xi)\right]  ^{-3/5}.
\end{align}
This means that turbulence cannot be neglected. For uplink, one can adopt Yura's model~\cite{Yura73} and decompose the long-term spot size of Eq.~\eqref{ltspot} as follows
\begin{equation}
w_{\text{st}}^{2}\simeq w_{h}^{2}+2\left(  \frac{\lambda h}{\pi\rho
_{0}^{\uparrow}}\right)  ^{2}\Psi,~~\sigma_{\text{TB}}^{2}\simeq2\left(
\frac{\lambda h}{\pi\rho_{0}^{\uparrow}}\right)  ^{2}(1-\Psi),
\end{equation}
where 
\begin{equation}
  \Psi :=[1-0.33 (  \rho_{0}^{\uparrow}/w_{0})^{1/3}]^2.  
\end{equation}
For a receiver with radius $a_R$, we consider the uplink short-term
transmissivity
\begin{equation}
\eta_{\text{st}}(h)=1-e^{-2a_{R}^{2}/w_{\text{st}}^{2}} \simeq\frac{2a_{R}^{2}%
}{w_{\text{st}}^{2}},
\label{shortTERMtransmissivity}
\end{equation}
and the uplink long-term transmissivity 
\begin{equation}
\eta_{\text{lt}}(h)=1-e^{-2a_{R}^{2}/w_{\text{lt}}^{2}} \simeq\frac{2a_{R}^{2}%
}{w_{\text{lt}}^{2}}.
\label{longTERMtransmissivity}
\end{equation}

If the detector is fast enough, it is able to track the fast dynamics of turbulence. In that case, the short-term transmissivity $\eta_{\text{st}}$ provides an upper bound for the uplink performance (however, this will be non-tight, because it will ignore the effect of beam wandering due to $\sigma_{\text{TB}}^{2}$). If the detector is slow, we need to compute the bound from the long-term transmissivity $\eta_{\text{lt}}$ (where the effect of beam wandering is averaged and further contributes to the beam widening)~\cite[Sec.~II.F]{freespace}. 

We therefore consider the total uplink transmissivity for a satellite at altitude $h$ (zenith position)
\begin{equation}
    \eta(h) = \eta_{\mr{eff}}\eta_{\mr{atm}}(h)\eta_{\mr{fs}}(h),
\end{equation}
where $\eta_{\text{fs}}(h)=\eta_{\text{st}}(h)$ or $\eta_{\text{lt}}(h)$, depending on the detector regime (fast or slow). Note that, for $h> 3 \times 10^4$~m, we can approximate $\eta_{\mr{atm}}(h)\simeq\eta_{\mr{atm}}^{\mr{zen}} \simeq 0.967$ at $\lambda=800$~nm~\cite{PirandolaSAT}.

We then compute $\gamma(h):=\gamma[\eta(h)]$ where the expression of $\gamma$ depends on the QKD implementation, i.e., single photon ($\gamma = \eta$) or attenuated source with intensity $\mu$ ($\gamma = 1 - e^{-\eta \mu}$). Finally, we impose $\gamma(h) > \Gamma$ where $\Gamma=\Gamma_2$ or $\Gamma_3$ is defined in Eqs.~(11) and~(12) of the main text, depending on the number of MUBs. Thus, we bound the maximum altitude $h_{\mr{max}}$ at which QKD with the satellite is possible (given the implementation parameters). The results are shown in Table~\ref{tableFSup}. We can see how the distance limits are reduced by one order of magnitude with respect to the downlink.
\begin{table}[b]
\centering
\begin{tabular}{|c|c|c|c|c|}
\hline
\multirow{2}{*}{} 
& \multicolumn{2}{c|}{2-MUB} 
& \multicolumn{2}{c|}{3-MUB} \\ 
\cline{2-5}
& 1 photon & $\mu=0.3$ & 1 photon & $\mu=0.3$ \\ 
\hline
Fast 
& $1.78 \times10^5$ & $0.97 \times10^5$ & $2.53 \times10^5$ & $1.38 \times10^5$ \\ 
\hline
Slow
&  $1.48 \times10^5$ & $0.81 \times10^5$ & $2.10 \times10^5$& $1.15 \times10^5$ \\ 
\hline
\end{tabular}
\caption{Bounds on the uplink free-space distance in km, for the various protocols assuming fast and slow detection. Parameters are chosen as in Table~\ref{tabDOWN}. Turbulence follows the  H-V$_{5/7}$ model (nighttime and low winds).}
\label{tableFSup}
\end{table}


\section{Zero-capacity thresholds for repeater chains}\label{app:rep}
For a qubit Pauli channel, the upper bound $\Phi(\mathcal{P})$ in Eq.~\eqref{strettoPAULI} comes from the relative entropy of entanglement $E_R$ (REE)~\cite{REE1,REE2} computed over its Choi state $\sigma_{\mathcal{P}}$. If we have a chain of $N$ intermediate repeaters between Alice and Bob, we have a total of $N+1$ Pauli channels $\mathcal{P}_i$ on the chain. According to Ref.~\cite[Th.~3]{repeaterCAP}, we have that the end-to-end capacity between Alice and Bob satisfies
\begin{equation}
    \mathcal{C}_2(\{\mathcal{P}_i\}) \le \min_i E_R(\sigma_{\mathcal{P}_i})=\min_i \Phi(\mathcal{P}_i).
\end{equation}
For channel $\mathcal{P}_i$ with probabilities $\{p_k^i\}$, let us define $p_{\max}^i:=\max_k{\{ p_k^i\}}$. Then, we take the minimum along the chain $p_{\max}^{\min}:=\min_i p_{\max}^i$, and we can write
\begin{align}
&\mathcal{C}_2(\{\mathcal{P}_i\}) \le  
\begin{cases}
1-H_2(p_{\max}^{\min}), & p_{\max}^{\min}\ge \tfrac12,\\
0, & \text{otherwise}.
\end{cases}
\label{strettoPAULI2}
\end{align}
This result means that $p_{\max}^{\min} \le 1/2$ certainly implies $\mathcal{C}_2(\{\mathcal{P}_i\})=0$. Unfortunately, the opposite cannot be proven using the simple approach of the repeaterless case. The problem is that the Bell detections performed at the repeaters (necessary to extend entanglement to the end points) will gradually decrease the fidelity if the states distributed over the links are not perfect Bell pairs. 

\FloatBarrier


\end{document}